\documentclass[conference]{IEEEtran}
\IEEEoverridecommandlockouts
\usepackage{algorithm}
\usepackage{algpseudocode}
\usepackage{amsmath}
\usepackage{array}
\usepackage{xcolor}
\usepackage[hidelinks,hyperfootnotes=false]{hyperref}
\hypersetup{draft}
\usepackage{comment}
\usepackage{cite}
\usepackage{booktabs}
\usepackage{subcaption}

\ifCLASSINFOpdf
  \usepackage[pdftex]{graphicx}
\def\BibTeX{{\rm B\kern-.05em{\sc i\kern-.025em b}\kern-.08em
    T\kern-.1667em\lower.7ex\hbox{E}\kern-.125emX}}
    
\begin{document}

\title{LLM-Guided Cyber-Physical Attack Path Identification in Smart Grids
\thanks{This work was supported by NSF Award 2220346.}}

\author{
    Amr~Elshazly$^*$, Jesse Roberts$^*$, Rachad Atat$^\dagger$, and Muhammad~Ismail$^*$ \\
    \small $^*$CEROC \& Department of Computer Science, Tennessee Tech University, Cookeville, TN, USA \\
    \small $^\dagger$Department of Computer Science, Lebanese American University, Beirut, Lebanon \\
    \small Emails: $^*$\{aaelshazly42, jtroberts, mismail\}@tntech.edu, $^\dagger$rachad.atat@lau.edu.lb
}

\maketitle

\begingroup
\renewcommand{\thefootnote}{}
\footnotetext{%
\copyright\ 2026 IEEE. Personal use of this material is permitted.
Permission from IEEE must be obtained for all other uses,
in any current or future media, including reprinting/republishing
this material for advertising or promotional purposes,
creating new collective works, for resale or redistribution
to servers or lists, or reuse of any copyrighted component
of this work in other works.%
}
\endgroup

\begin{abstract}
Identifying high-impact attack paths in cyber-physical power systems is critical for improving grid resilience. Existing methods typically rely on either structural topology analysis or simulation-based cascading-failure models. Topology-based approaches capture connectivity but lack adversarial reasoning, while simulation-driven methods are primarily suited for offline analysis and are less effective for rapid evaluation of evolving attack scenarios. As a result, identifying realistic attack paths that jointly capture topology, threat semantics, and physical impact remains challenging. This paper presents a physics-grounded framework for identifying cyber-physical attack paths using large language models (LLMs), with cascading-failure-derived impact rankings used as physics-grounded references. The framework integrates system topology, LLM-guided STRIDE and MITRE ATT\&CK threat modeling, and three baseline strategies for comparison: topology-driven, threat-driven, and fusion-based. Evaluation is performed using (i) critical-node overlap between LLM-inferred top-$K$ rankings and cascading-failure-based ground truth; (ii) node-level similarity between LLM-inferred and reference paths; and (iii) path-level alignment capturing agreement in complete attack sequences. Results show that structured LLM-based threat modeling substantially improves alignment with physics-grounded references. In the cyber layer, critical-node overlap@3 increases from approximately 33\% to as high as 100\%, while node-level similarity reaches up to 65\% and path-level alignment reaches up to 48\%. In addition, the proposed framework reduces runtime by approximately 75--85\% compared to the fusion-based baseline in the IEEE~14-bus system. These results suggest that LLM-guided attack path identification is a promising approach for efficient and impact-aware cyber-physical security analysis.
\end{abstract}

\begin{IEEEkeywords}
Cyber-physical systems, attack path analysis, large language models, cascading failure, threat modeling
\end{IEEEkeywords}

\section{Introduction}

Cyber-Physical Power Systems (CPPS), such as smart grids, tightly integrate communication networks with physical power infrastructure~\cite{4519604, elshazly2024reinforcement}, expanding the attack surface and allowing cyber disruptions to propagate into cascading failures and large-scale blackouts~\cite{6032699, buldyrev2010catastrophic}. Frameworks such as STRIDE~\cite{shostack2014threat} and MITRE ATT\&CK~\cite{strom2018mitre} provide structured threat representations, but do not capture how cyber actions translate into physical impacts. Existing approaches rely on either topology analysis or simulation-based cascading-failure models~\cite{5477189, 6032699}. Topology-based methods lack adversarial reasoning, while simulation-driven approaches are primarily suited to offline analysis. As a result, identifying realistic attack paths that jointly account for topology, threat semantics, and physical impact remains challenging. Recent advances in LLMs have demonstrated strong structured reasoning capabilities~\cite{brown2020language}, motivating their use for impact-aware attack path identification.

This paper proposes a physics-grounded framework in which LLMs identify cyber-physical attack paths using system topology and structured threat intelligence from STRIDE~\cite{shostack2014threat} and MITRE ATT\&CK~\cite{strom2018mitre}, while evaluation is performed against cascading-failure-based references~\cite{10066436}. Attack paths are compared against three baselines: topology-driven, threat-driven, and fusion-based, using three metrics: critical-node overlap, node-level similarity, and path-level alignment capturing agreement in complete attack sequences. Our main contributions include:

\begin{itemize}
    \item An LLM-based framework for identifying cyber-physical attack paths in power systems using system topology and structured threat intelligence derived from STRIDE and MITRE ATT\&CK.
    
    \item A physics-grounded evaluation framework based on cascading-failure simulations, including topology-driven, threat-driven, and fusion-based baselines, and quantitative metrics such as critical-node overlap, node-level similarity, and path-level alignment. Physics-grounded refers to evaluation against impact rankings obtained from cascading-failure simulations, rather than to direct embedding of power-flow equations inside the LLM prompt.
    
    \item Experimental validation on the IEEE 14-bus cyber-physical test system demonstrating improved alignment between LLM-identified paths and impact-aware baselines, with significantly reduced runtime compared with the fusion-based baseline.
\end{itemize}

The remainder of this paper is organized as follows. Section~\ref{sec:related_work} reviews related work, Section~\ref{sec:methodology} presents the proposed framework, Section~\ref{sec:results} reports the results, Section~\ref{sec:discussion} discusses the findings, and Section~\ref{sec:conclusion} concludes the paper.


\section{Related Work}
\label{sec:related_work}

Cyber-physical attack path analysis has been extensively studied from multiple perspectives, including graph-based modeling, threat-driven reasoning, physics-based analysis, and recent LLM-assisted approaches. Most prior work originates from the cybersecurity and general cyber-physical systems (CPS) domains, with a subset specifically focused on CPPS.

\vspace{-1mm}

\subsection{Graph-Based Attack Modeling}

Graph-based methods, such as attack graphs and attack trees, model multi-step attacks by enumerating feasible sequences~\cite{sheyner2002attackgraph}. Recent approaches incorporate vulnerability data and threat intelligence, including knowledge graph-based frameworks~\cite{attacKG2021, binyamini2020attackmodeling}. While effective at capturing structural feasibility, these methods are primarily connectivity-driven and do not account for adversarial reasoning or physical impact.

\vspace{-1mm}

\subsection{Threat Modeling and Semantic Reasoning}

Frameworks such as STRIDE~\cite{shostack2014threat} and MITRE ATT\&CK~\cite{strom2018mitre} provide structured representations of adversarial behavior and have been used to guide attack reasoning~\cite{husari2017ttpdriven, zhang2019attackchain}. These approaches are generally developed for cybersecurity or general CPS settings and are largely limited to the cyber domain, without modeling physical system dynamics and impacts.

\subsection{Physics-Based Analysis}

Cascading-failure models are widely used in interdependent networks and power systems to analyze failure propagation and identify critical components~\cite{buldyrev2010catastrophic}. In CPPS contexts, these models provide insight into system vulnerability and impact, but focus on identifying critical nodes rather than modeling adversarial attack strategies or identifying high-impact cyber-physical attack paths.


\subsection{Data-Driven Approaches}

Data-driven methods infer attack paths from observed events or system behavior, such as SAGE~\cite{sage2021attackgraph} and recent survey literature on attack path analysis~\cite{attackgraphsurvey2023}. While scalable, they are limited to cyber observations and do not integrate structured threat knowledge or physical system dynamics and impacts.

\subsection{LLMs in Cyber-Physical Security}

Recent work has applied LLMs to cybersecurity tasks such as threat detection, vulnerability analysis, and attack prediction~\cite{zhang2025llms, li2025texts}, with extensions to CPS for anomaly detection and threat modeling~\cite{diaf2024beyond, feng2023llminvariant, 11143218}. However, these efforts focus on isolated tasks and do not address topology-constrained, multi-step cyber-to-physical attack path identification in power systems or integrate threat semantics with physics-grounded evaluation.

\subsection{Research Gap}
Prior works address key aspects of cyber-physical security, including graph-based attack modeling~\cite{sheyner2002attackgraph, attacKG2021, binyamini2020attackmodeling}, threat modeling~\cite{husari2017ttpdriven, zhang2019attackchain}, and cascading-failure-based impact analysis~\cite{buldyrev2010catastrophic}, but these components are typically studied separately. No prior study uses LLMs to generate multi-step cyber-to-physical attack paths in CPPS by jointly combining topology-aware reasoning, STRIDE- and MITRE-based threat semantics, and evaluation against cascading-failure-derived physical-impact references. This paper introduces an LLM-based framework for multi-step cyber-to-physical attack path identification with physics-grounded evaluation.

\section{Methodology}
\label{sec:methodology}

\begin{figure*}[t]
    \centering
    \includegraphics[width=\textwidth, height=6cm]{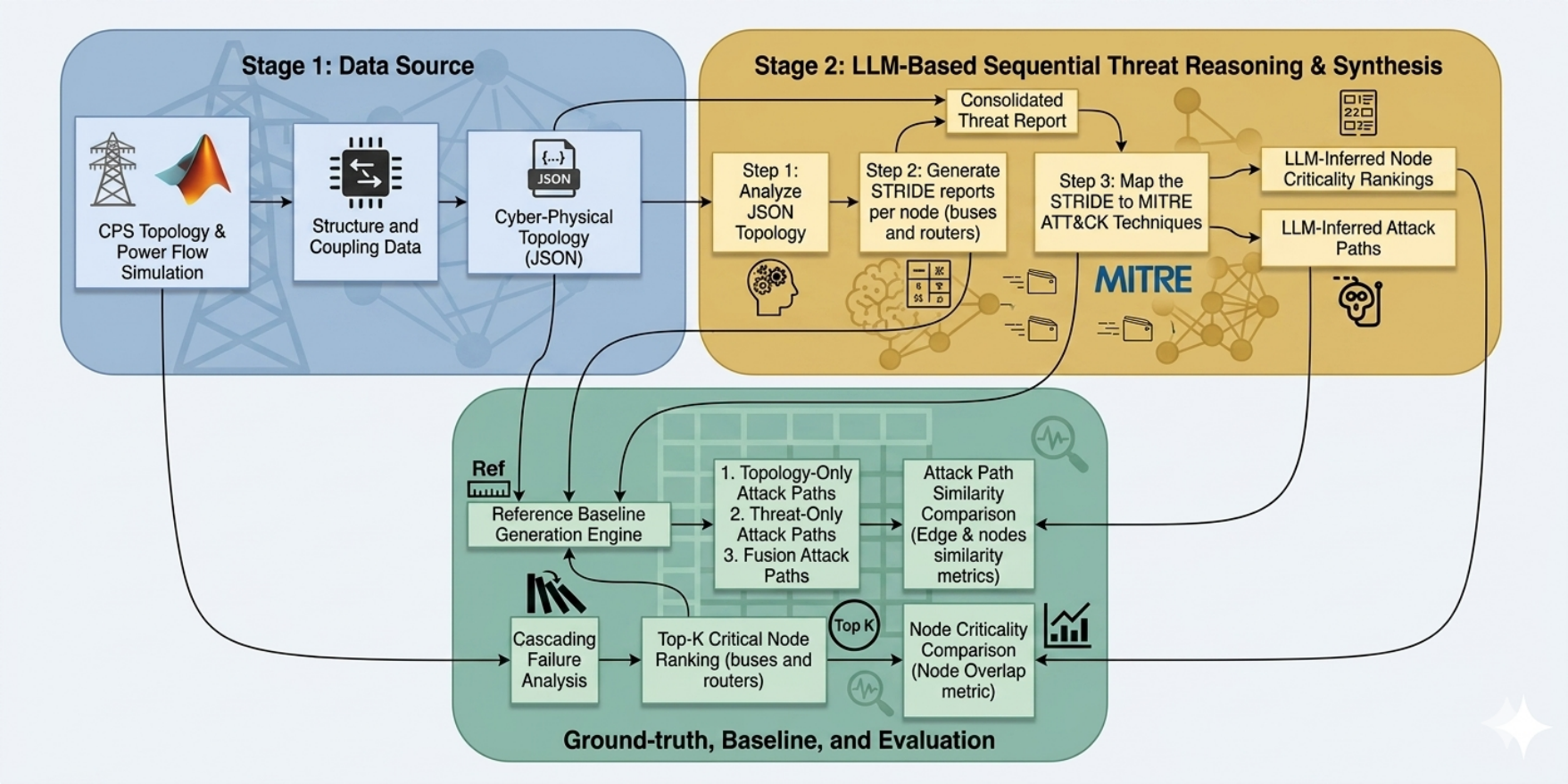}
    \caption{Proposed physics-grounded LLM framework for cyber-physical attack path identification and evaluation. The upper branch shows LLM-based inference of node rankings and attack paths, while the lower branch provides physics-grounded references from cascading-failure analysis~\cite{10066436} and baseline strategies for comparison.}
    \label{fig:framework}
\end{figure*}

This section presents the proposed framework for identifying cyber-physical attack paths using LLMs and evaluating them against physics-based impact rankings derived from cascading-failure analysis, as shown in Fig.~\ref{fig:framework}. The framework consists of two main components: (i) LLM-based attack path identification, and (ii) physics-based ground-truth construction. Cascading-failure analysis is used only to construct physics-grounded reference rankings and baseline paths for evaluation. These impact rankings are not provided to the LLM during attack-path generation, which preserves a separation between LLM inference and physics-based validation.

\subsection{Stage 1: Physics-Grounded Data and Impact Modeling}

We represent the cyber-physical power system (CPPS) as a JSON-encoded graph that serves as input for LLM-based reasoning, capturing both the cyber and physical layers along with their interdependencies.
\begin{itemize}
    \item \textbf{Physical layer:} Buses with functional roles (e.g., generators and loads) and their electrical connectivity.
    \item \textbf{Cyber layer:} Routers representing communication nodes and their connectivity.
    \item \textbf{Inter-layer coupling:} Router-bus associations enabling cyber-physical interactions.
\end{itemize}

Formally, the topology is defined by the physical node set $\mathcal{V}_p$ (buses) with $|\mathcal{V}_p| = N_B$, the cyber node set $\mathcal{V}_c$ (routers) with $|\mathcal{V}_c| = N_c$~\cite{ieee14bus}, the physical edge set $\mathcal{E}_p$ (transmission branches), the cyber edge set $\mathcal{E}_c$ (communication links), and the inter-layer coupling set $\mathcal{E}_{cp} \subseteq \mathcal{V}_p \times \mathcal{V}_c$ (router-bus associations), as defined in~\cite{10066436}. Physics-grounded impact scores $I(n)$ are derived from cascading-failure simulations using MATPOWER, where each node is treated as a failure seed and the resulting system degradation is quantified via load loss and connectivity disruption. These scores reflect each node's criticality and are used for baseline construction and evaluation, as detailed in Section~\ref{subsec:ground_truth}.

\subsection{Stage 2: LLM-Based Threat Reasoning and Attack Path Identification}

In this stage, the LLM acts as a reasoning engine that transforms the JSON-encoded CPPS topology into threat-aware attack paths. Reasoning focuses on the cyber layer and cyber-physical coupling to target the physical domain. Through structured prompting, the model processes the topology to generate constrained, multi-step outputs: (i) STRIDE threat reports and mapped MITRE ATT\&CK techniques, (ii) node criticality rankings, and (iii) validated cyber-physical attack paths, as illustrated by the steps in Fig.~\ref{fig:framework}. The reasoning pipeline proceeds in three logical steps. First, the LLM interprets the JSON topology to identify cyber nodes, their connectivity, and their coupling relationships with physical components. Based on this representation, the model generates STRIDE-based threat reports describing vulnerabilities and attack scenarios associated with cyber nodes and communication links.

Second, the LLM maps each STRIDE threat to corresponding MITRE ATT\&CK techniques, translating abstract vulnerabilities into concrete adversarial actions. Each threat is mapped to one or more MITRE ATT\&CK techniques along with their corresponding tactics (e.g., Initial Access and Lateral Movement), enabling representation of realistic attacker behavior grounded in threat intelligence.

\begin{figure*}[t]
    \centering
    \includegraphics[width=\textwidth, height=6.4cm]{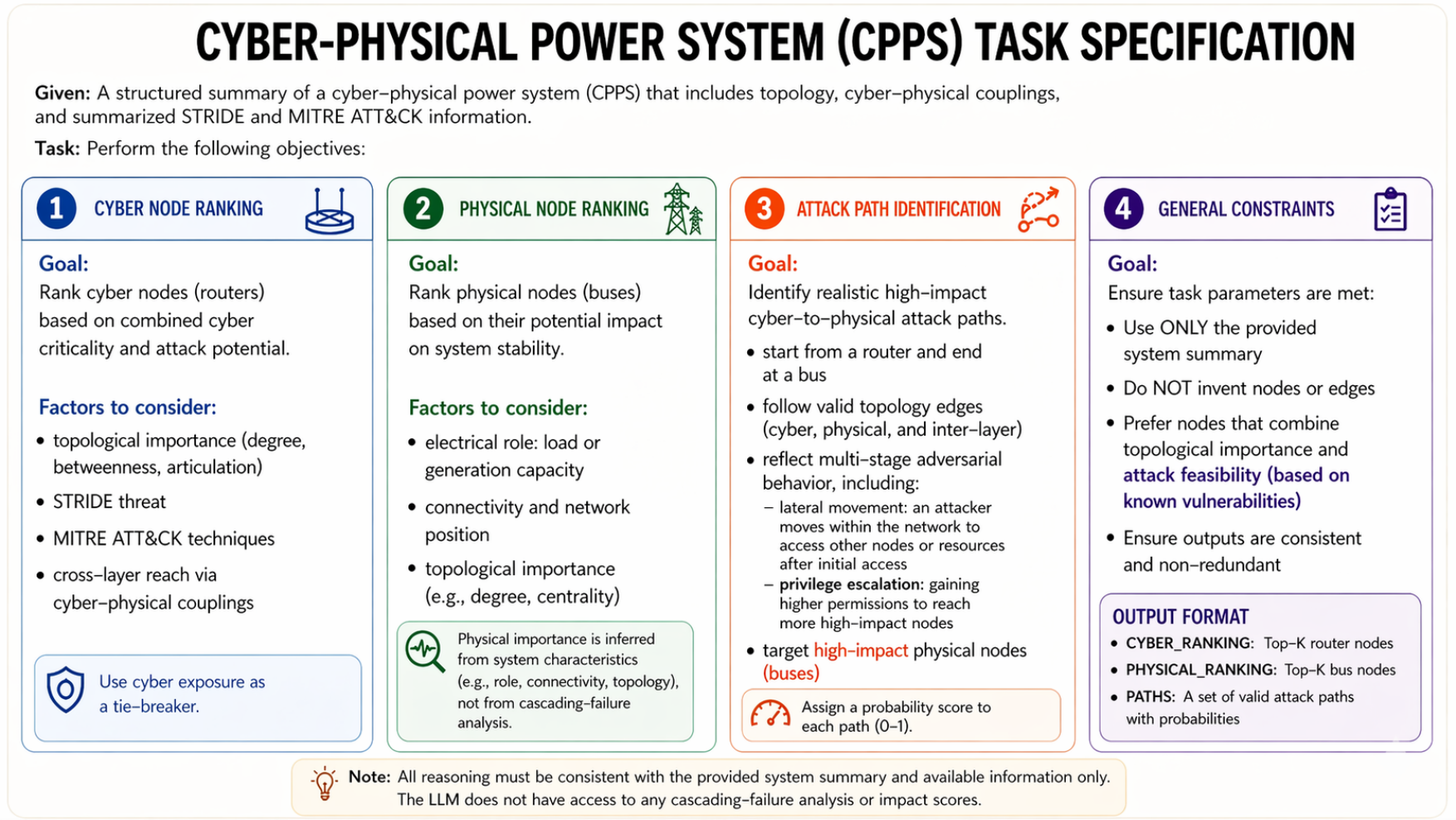}
    \caption{Summary of the LLM prompt used to guide threat reasoning and cyber-physical high-impact attack path identification. The prompt encodes system topology, STRIDE-based threat modeling, MITRE ATT\&CK mapping, and task-specific constraints to produce structured outputs, including node impact rankings and high-impact attack paths.}
    \label{fig:prompt}
\end{figure*}

Finally, the LLM leverages threat semantics and cyber connectivity to generate structured outputs, including top-$K$ node criticality rankings and multi-step cyber-physical attack paths. Node importance reflects (i) cyber risk derived from STRIDE/MITRE analysis and (ii) structural relevance to critical physical components based on topology and coupling. These rankings are compared against cascading-failure-based references~\cite{10066436}, as detailed in Section~\ref{subsec:ground_truth}. The generated attack paths originate in the cyber layer and propagate toward high-impact physical nodes via cyber-physical coupling. The structured prompt (Fig.~\ref{fig:prompt}) enforces explicit reasoning steps and a fixed input-output schema to ensure consistent and reproducible outputs. All experiments use identical prompts and fixed decoding parameters (temperature = 0.5, top-$p$ = 0.9, maximum tokens = 800), ensuring that variability arises from controlled sampling rather than prompt changes. This design enables systematic and fair comparison across models and baselines. All results are obtained through repeated LLM inference (rather than single-shot prompting) under fixed prompts and decoding settings, and are averaged over 1,000 independent runs, as detailed in Section~\ref{sec:results}.

\subsection{Ground Truth and Baseline Construction}
\label{subsec:ground_truth}

This subsection describes the components used to establish physics-grounded reference data and construct baseline strategies for comparison. We first derive impact-based ground-truth rankings via cascading-failure simulations, followed by the development of baseline path-generation methods capturing different levels of guidance in attack construction.

\subsubsection{Impact Modeling (Ground Truth)}

To derive physics-grounded reference rankings, we perform cascading-failure simulations~\cite{10066436} in MATLAB using the same cyber-physical topology. This ensures consistency between the structured system representation and the physics-based analysis. Node impact, defined as the system-level degradation resulting from the failure of a node across both cyber and physical layers, is assessed through node-wise disturbance analysis, where each cyber or physical node is treated as a failure seed and the resulting cascading effects are measured across both layers.
\begin{equation}
I(n)=w_1\,\mathrm{PoDN}(n)+w_2\,\mathrm{DamP}(n)+w_3\,\mathrm{DamC}(n),
\label{equ:1}
\end{equation}
where $\mathrm{PoDN}(n)$~\cite{10066436} denotes the load-shedding ratio, $\mathrm{DamP}(n)$~\cite{10066436} denotes the normalized loss of physical-network efficiency, and $\mathrm{DamC}(n)$ denotes the normalized loss of cyber-network efficiency. These metrics quantify the degradation in physical and cyber network performance following the failure of node $n$. Running the simulations for all nodes produces separate impact rankings for buses and routers, from which the top-$K$ nodes in each layer are selected as the physics-grounded ground truth for evaluating the LLM-based rankings.

\subsubsection{Baseline Path Identification}

To enable quantitative evaluation, we construct three baseline cyber-physical attack path identification strategies: topology-driven, threat-driven, and fusion-based approaches, each capturing a different level of guidance in attack path construction.

\paragraph{Topology-Driven Baseline}
Attack paths are generated by traversing valid cyber and physical connections based solely on structural reachability, without incorporating any threat intelligence or impact information. Node selection is not guided by any scoring or prioritization mechanism. To ensure a clean separation from informed strategies, no threat- or impact-based criteria are used during path construction. As a result, the topology-driven baseline approximates a structure-constrained random exploration of the network governed by topological connectivity.

\paragraph{Threat-Driven Baseline}
Attack paths are guided by cyber risk scores $R(n)$ derived from STRIDE and MITRE ATT\&CK reports. For each node, STRIDE-based threats are extracted and weighted according to category severity. Specifically, Elevation of Privilege is assigned the highest weight (1.0), followed by Spoofing (0.8), Tampering (0.7), Denial of Service (0.5), Information Disclosure (0.3), and Repudiation (0.2). In parallel, each MITRE ATT\&CK technique associated with a node contributes a fixed additive weight. The raw risk score $R(n)$ is computed as the sum of the STRIDE and MITRE contributions, and the resulting values are normalized using min-max scaling across all nodes. Nodes with higher normalized scores $\hat{R}(n)$ are prioritized during traversal, enabling threat-aware path construction. However, this approach does not incorporate physical impact information, limiting its ability to guide paths toward high-impact physical targets.

\paragraph{Fusion-Based Baseline}
The fusion-based approach integrates cyber risk and physical impact to approximate a realistic attacker strategy. Each node $n$ is assigned a risk score $R(n)$, derived from STRIDE and MITRE ATT\&CK analysis, and an impact score $I(n)$, obtained from cascading-failure simulations according to~(\ref{equ:1}). To ensure comparability, both scores are normalized via min-max scaling to $\hat{R}(n)$ and $\hat{I}(n)$, respectively. The normalized scores are then combined:

\begin{equation}
F(n) = \lambda \hat{R}(n) + (1 - \lambda)\hat{I}(n),
\end{equation}

\noindent where $\lambda \in [0,1]$ controls the trade-off between cyber risk and physical impact. Candidate entry points are selected from a top-$K$ pool of high-risk routers based on $\hat{R}(n)$, reflecting the attacker's preference for targeting vulnerable cyber infrastructure. Here, $K$ denotes the number of candidate entry-point routers, while $K_t$ denotes the number of critical target buses used to define high-impact transition points. In the fusion setting, the source cyber node is sampled probabilistically using a softmax distribution over the fused scores:

\begin{equation}
P_s(n) = \frac{\exp(\beta F(n))}{\sum_{i \in \mathcal{R}_K} \exp(\beta F(i))}, 
\quad n \in \mathcal{R}_K
\label{equ: probability}
\end{equation}

\noindent where $n \in \mathcal{R}_K$ denotes a candidate cyber node selected from the top-$K$ router set, and $\beta > 0$ is a temperature parameter controlling the sharpness of the distribution, with higher values more strongly favoring high-scoring nodes. This sampling strategy prioritizes high-impact cyber entry points while preserving exploration of diverse attack trajectories. Attack paths are constructed through a transition-aware cyber walk. For a current node $u$, each neighboring node $v$ is assigned a transition score:

\begin{equation}
S(u,v) = \alpha F(v) + (1 - \alpha)\Phi(v),
\end{equation}

\noindent where $\alpha \in [0,1]$ balances immediate node importance and future potential. The first term $F(v)$ represents the current importance of node $v$, while the second term $\Phi(v)$ captures its future utility. The look-ahead term is defined as:

\begin{equation}
\Phi(v) = \max_{w \in \mathcal{N}_c(v)} F(w),
\end{equation}

\noindent where $\mathcal{N}_c(v)$ denotes the set of cyber-layer neighboring nodes of $v$. This term identifies the most critical node reachable from $v$ in one additional step, capturing the best downstream opportunity after transitioning to $v$. After computing $S(u,v)$ for all neighbors of $u$, the scores are converted into a probability distribution using a softmax function similar to~(\ref{equ: probability}). The next node is then sampled from this distribution.

During traversal, the algorithm monitors the visited cyber nodes to identify potential cyber-physical transition points. A node $v$ is added to the transition candidate set $\mathcal{T}$ if it is coupled to a physical bus $b$ such that $b \in \mathcal{B}_{K_t}$ (or is a direct neighbor of one). This condition ensures that the selected cyber node is associated with a high-impact physical target, while the traversal itself remains confined to the cyber layer. Once such a node is reached, the cyber traversal terminates to avoid unnecessary exploration. Any cycles within the visited sequence are removed to obtain a simple cyber path $\pi_c$. The final attack path is then represented as $\pi = \pi_c \rightarrow b^*$, where $b^*$ denotes the high-impact physical target associated with the terminal cyber node, as formalized in Algorithm~\ref{alg:fusion}.

\begin{algorithm}[!b]
\small
\caption{Fusion-Based Baseline for Cyber-Physical Attack Path Generation}
\label{alg:fusion}
\begin{algorithmic}[1]

\Require Cyber graph $(\mathcal{V}_c, \mathcal{E}_c)$, physical graph $(\mathcal{V}_p, \mathcal{E}_p)$, coupling set $\mathcal{E}_{cp}$
\Require STRIDE/MITRE reports, cascading-failure impact scores $I(n)$
\Require $K$: number of candidate entry routers; $K_t$: number of critical target buses
\Require $\lambda \in [0,1]$: cyber/impact trade-off; $\alpha \in [0,1]$: local/future balance
\Require $\beta > 0$: softmax temperature; MAXSTEPS: maximum cyber walk length
\Ensure Attack path $\pi = \pi_c \rightarrow b^*$, where $\pi_c$ is the cyber path and $b^*$ is the associated physical target

\Statex \textit{\textbf{Precomputation}}
\ForAll{$n \in \mathcal{V}_c$}
    \State Compute normalized scores $\hat{R}(n)$ (cyber risk) and $\hat{I}(n)$ (physical impact)
    \State $F(n) \gets \lambda \hat{R}(n) + (1 - \lambda)\hat{I}(n)$
\EndFor
\State $\mathcal{R}_K \gets$ Top-$K$ routers ranked by $\hat{R}(n)$ \hfill \textit{// candidate entry points}
\State $\mathcal{B}_{K_t} \gets$ Top-$K_t$ buses ranked by $\hat{I}(n)$ \hfill \textit{// high-impact physical targets}

\Statex \textit{\textbf{Source Selection}}
\State Sample source $s \in \mathcal{R}_K$ with $P_s(n) \propto \exp(\beta F(n))$
\State Initialize $\pi_c \gets [s]$, $\mathcal{T} \gets \emptyset$

\Statex \textit{\textbf{Transition-Aware Cyber Walk}}
\While{$|\pi_c| <$ MAXSTEPS}
    \State $u \gets$ last node in $\pi_c$
    \ForAll{$v \in \mathcal{N}_c(u)$}
        \State $\Phi(v) \gets \max_{w \in \mathcal{N}_c(v)} F(w)$
        \State $S(u,v) \gets \alpha F(v) + (1-\alpha)\Phi(v)$
    \EndFor
    \State Sample $v^* \in \mathcal{N}_c(u)$ with $P(v^* \mid u) \propto \exp(\beta S(u,v^*))$
    \State Append $v^*$ to $\pi_c$
    \If{$\exists b \in \mathcal{B}_{K_t} \text{ such that } (v^*, b) \in \mathcal{E}_{cp}$}
        \State $\mathcal{T} \gets \mathcal{T} \cup \{v^*\}$
        \State \textbf{break}
    \EndIf
\EndWhile

\State Remove cycles from $\pi_c$

\Statex \textit{\textbf{Target Identification}}
\State Select $v^* \in \mathcal{T}$ such that 
$b^* = \arg\max_{b:\,(v^*,b)\in\mathcal{E}_{cp},\,b\in\mathcal{B}_{K_t}} \hat{I}(b)$
\State \Return $\pi = \pi_c \rightarrow b^*$

\end{algorithmic}
\end{algorithm}

\section{Results and Evaluation}
\label{sec:results}

This section evaluates the proposed framework for identifying cyber-physical attack paths. Experiments are conducted on a CPPS based on the IEEE 14-bus system~\cite{ieee14bus}, where the physical layer is derived from MATPOWER and a corresponding cyber layer of routers is mapped to buses to model cyber-physical coupling~\cite{buldyrev2010catastrophic, 9641756}. The same graph representation is used for both cascading-failure simulations and LLM reasoning to ensure consistency. LLM-based reasoning is implemented using instruction-tuned models, including LLaMA 3 (8B), Mistral-Nemo Instruct (2407), and Qwen2.5 (14B)~\cite{llama3modelcard, mistralnemo2024, qwen2.5}, which support structured multi-step reasoning~\cite{llm_reasoning_ref}. Results are averaged over 1,000 independent runs using fixed prompts and controlled decoding settings (temperature and top-$p$) to ensure reproducibility.

\subsection{Evaluation Metrics}
We evaluate the identified attack paths using three metrics:

\textit{Node-Level Overlap:} Overlap between LLM-inferred top-$K$ node rankings and cascading-failure ground truth, reported as overlap@$(3,5,7)$ for cyber and physical layers.

\textit{Path-Level Similarity:} Alignment between LLM-inferred and baseline paths, evaluated independently for each baseline. Each baseline defines its reference set of paths; therefore, similarity is computed separately with respect to each baseline. A generated path is counted as a match only if it exactly matches the corresponding reference path in both node sequence and traversal order. The reported score represents the percentage of matched paths, averaged across all generated paths and 1,000 runs. Scores are not additive across baselines.

\textit{Runtime:} Time to identify high-impact attack paths, compared against baseline references.

\subsection{Impact of Threat Modeling (STRIDE + MITRE)}

\begin{figure*}[t!]
    \centering
    \begin{subfigure}[b]{0.49\textwidth}
        \centering
        \includegraphics[width=\textwidth, height=5.1cm]{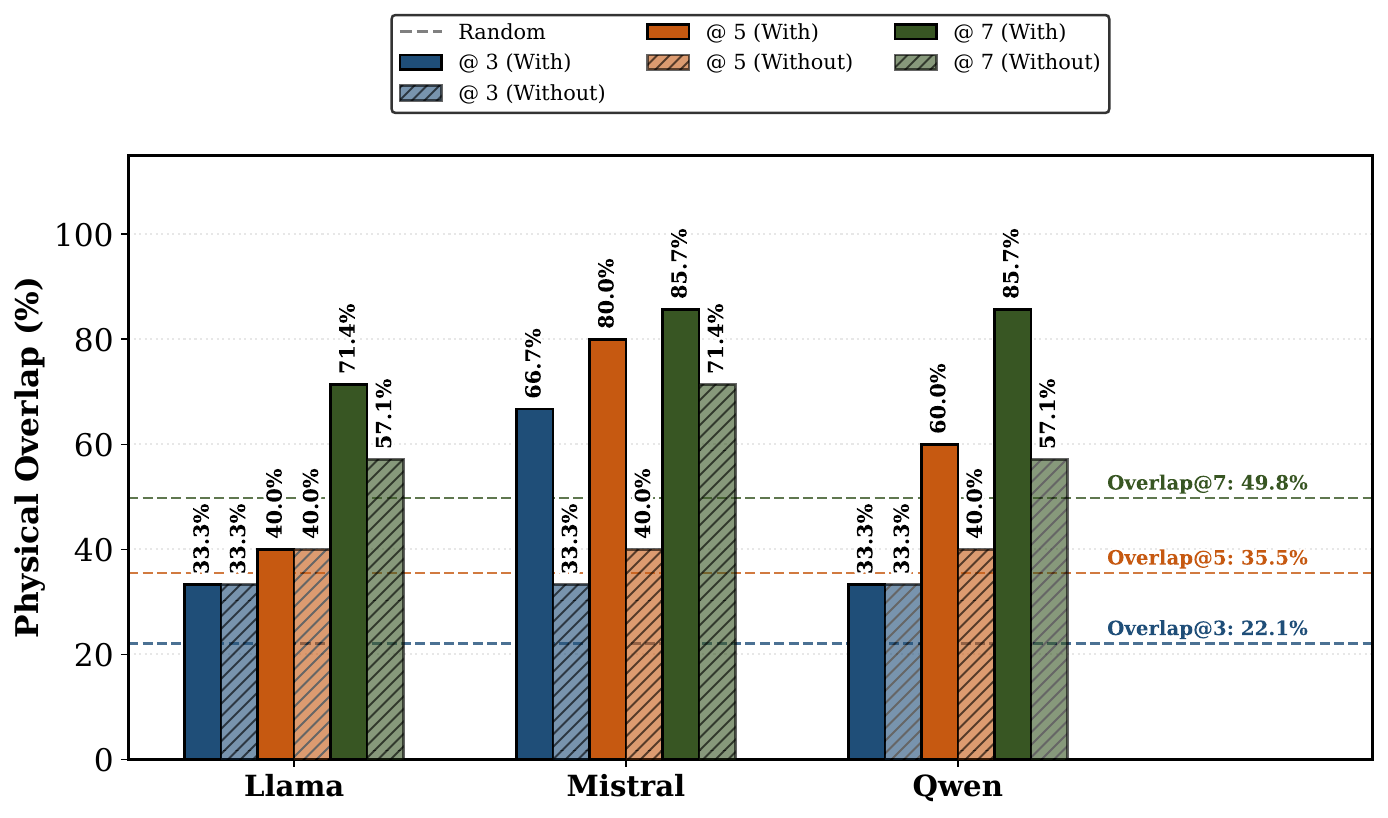}
        \caption{Physical critical node overlap comparison.}
        \label{fig:physical_overlap}
    \end{subfigure}
    \hfill
    \begin{subfigure}[b]{0.49\textwidth}
        \centering
        \includegraphics[width=\textwidth, height=5.1cm]{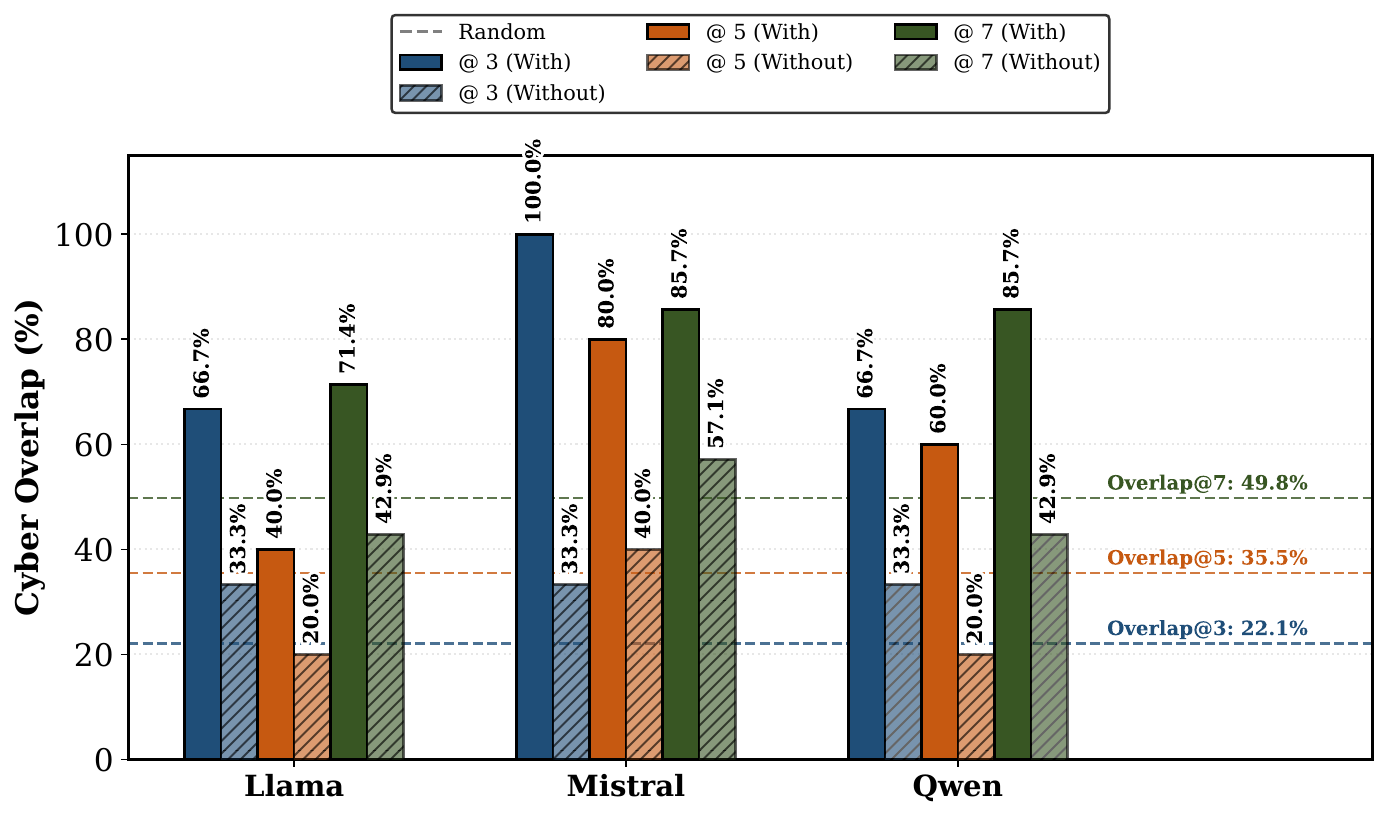}
        \caption{Cyber critical node overlap comparison.}
        \label{fig:cyber_overlap}
    \end{subfigure}
    
    \caption{Comparison of critical node overlap percentages (@3, @5, and @7) with and without STRIDE/MITRE reports across LLaMA, Mistral, and Qwen models, evaluated against cascading-failure-based ground truth. Dashed lines indicate random baseline (chance-level) overlap rates estimated via Monte Carlo simulation.}
    \label{fig:overlap_results}
\end{figure*}

Figs.~\ref{fig:physical_overlap} and~\ref{fig:cyber_overlap} compare node-level overlap scores generated with and without STRIDE/MITRE reports against cascading-failure ground truth. A random baseline, estimated via Monte Carlo simulation over 10,000 trials, provides a chance-level reference corresponding to topology-constrained random selection of nodes. In the physical layer, incorporating STRIDE and MITRE improves alignment, with overlap@7 reaching approximately 85\% for Mistral and Qwen, and coverage of high-impact buses exceeding 80\% at overlap@5. In the cyber layer, models initially show limited overlap ($\approx$30--35\%) without threat reports; integrating STRIDE and MITRE increases overlap@3 to 66--100\% depending on the model, with Mistral achieving full alignment. In both layers, results substantially exceed the random baseline, indicating that improvements reflect structured, threat-aware reasoning rather than chance.

\subsection{Node-Level Path Similarity}

Fig.~\ref{fig:node_similarity} compares node-level similarity between LLM-inferred and baseline paths. With STRIDE and MITRE, LLaMA and Mistral show stronger alignment with the fusion-based baseline (up to 60--65\%), while Qwen exhibits slightly lower but consistent performance. Without STRIDE and MITRE, all models shift toward topology-driven behavior with reduced alignment to fusion- and threat-driven baselines, highlighting the role of threat modeling in guiding impact-aware attack paths.

\subsection{Path-Level Similarity (Order-Sensitive)}

\begin{figure*}[t!]
    \centering
    \begin{subfigure}[b]{0.49\textwidth}
        \centering
        \includegraphics[width=\textwidth, height=4.5cm]{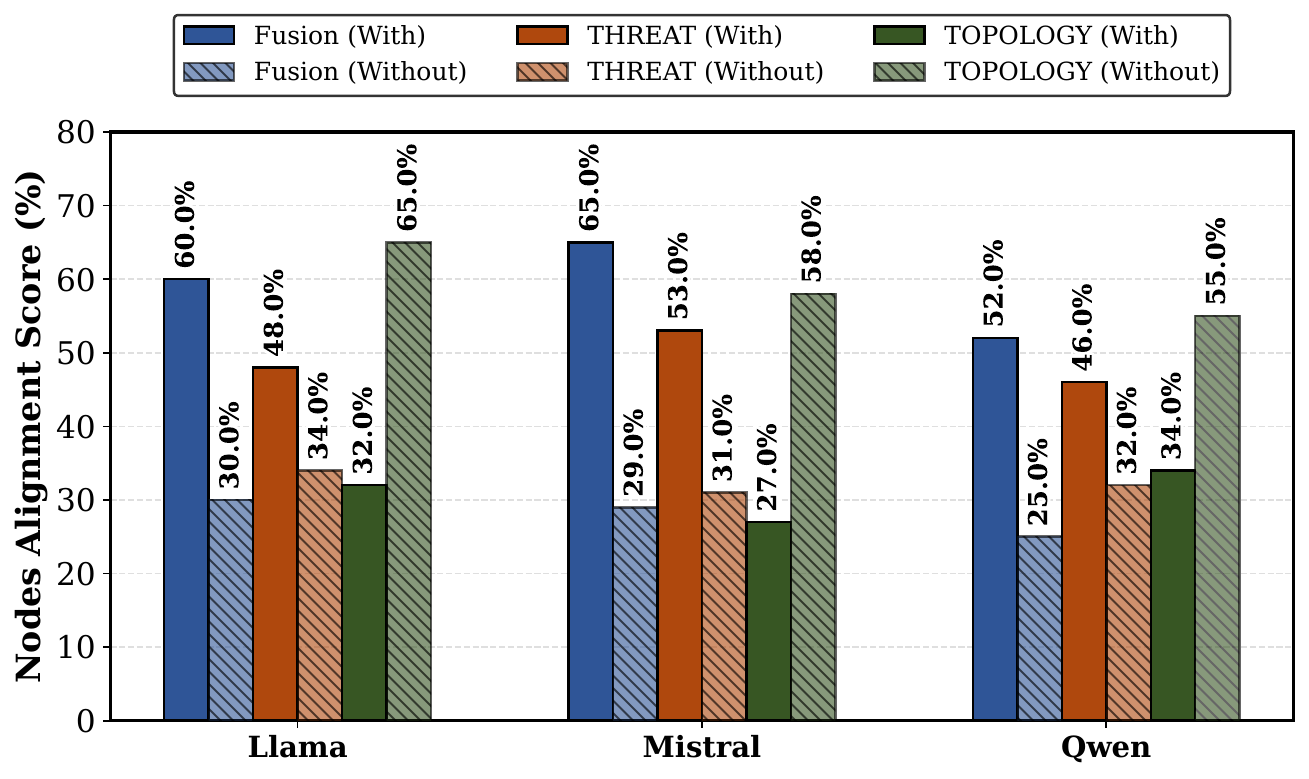}
        \caption{Node Alignment Score.}
        \label{fig:node_similarity}
    \end{subfigure}
    \hfill
    \begin{subfigure}[b]{0.49\textwidth}
        \centering
        \includegraphics[width=\textwidth, height=4.5cm]{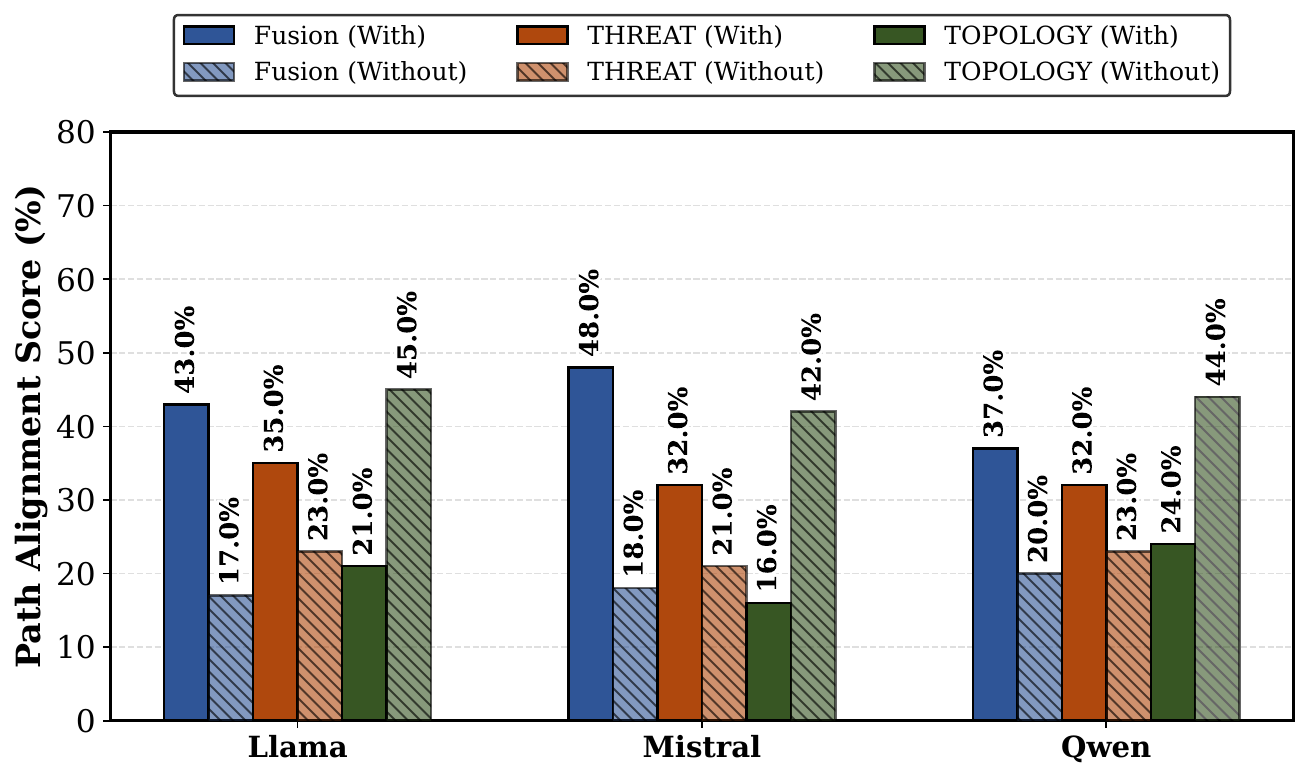}
        \caption{Full Path Alignment Score.}
        \label{fig:path_similarity}
    \end{subfigure}
    
    \caption{Alignment scores for node identification and full attack paths across LLaMA, Mistral, and Qwen. Scores are computed independently with respect to each baseline strategy and should not be interpreted as cumulative across baselines.}
    \label{fig:similarity_results}
\end{figure*}

Path-level similarity is reported as the percentage of generated paths that match reference paths, averaged across all generated paths and 1,000 independent runs. For example, a score of 48\% indicates that 48\% of generated paths fully match the corresponding fusion-based reference paths; it does not denote partial overlap within each path. Fig.~\ref{fig:path_similarity} evaluates both node selection and traversal order. Scores are lower than node-level metrics due to the stricter requirement of matching complete ordered paths. Mistral achieves the highest alignment (up to 48\%), followed by LLaMA, while Qwen shows lower scores. Without STRIDE and MITRE, similarity shifts toward topology-driven behavior, indicating structure-driven path generation. Overall, path-level alignment remains more challenging due to the need to capture full multi-step sequences. Similarity scores are computed independently with respect to each baseline strategy and should not be interpreted as cumulative across baselines.

\subsection{Runtime Analysis}
Table~\ref{tab:runtime} summarizes the average runtime per identified path. The fusion-based baseline requires 344--355 seconds, including both cascading-failure simulation ($\approx$180\,sec) for impact estimation and path generation (164--175\,sec). In contrast, the LLM-based approach does not require this offline simulation step. The LLM-based approach requires 40--90 seconds, comparable to the topology-driven baseline ($\approx$60\,sec), while reducing execution time by approximately 40--50\% compared to the threat-driven baseline and 75--85\% compared to the fusion-based baseline. These results demonstrate the efficiency of the proposed framework in a small-scale setting (IEEE 14-bus). Scalability to larger systems and the resulting impact on runtime remain topics for future work.

\begin{table}[t]
\centering
\caption{Average runtime per identified attack path.}
\label{tab:runtime}
\begin{tabular}{l c}
\hline
\textbf{Method} & \textbf{Runtime (per path)} \\
\hline
LLaMA 3 (8B)   & 40--70 sec \\
Mistral-Nemo   & 50--80 sec \\
Qwen2.5        & 60--90 sec \\
\hline
Topology-driven baseline  & 60--65 sec \\
Threat-driven baseline    & 110--117 sec \\
Fusion-based baseline     & 344--355 sec$^{\dagger}$ \\
\hline
\multicolumn{2}{l}{$^{\dagger}$Includes $\approx$180\,sec 
cascading-failure simulation.}
\end{tabular}
\end{table}

\section{Discussion}
\label{sec:discussion}

The results highlight several key observations regarding LLM-based cyber-physical attack reasoning.

Incorporating STRIDE and MITRE reports consistently improves alignment with high-impact nodes across both layers, as shown in Figs.~\ref{fig:physical_overlap} and~\ref{fig:cyber_overlap}. In the cyber layer, overlap@3 increases from about one-third without threat reports to 66--100\% with STRIDE/MITRE, while in the physical layer overlap@7 reaches up to 85\%. These threat-informed configurations exceed the random baseline by a wide margin, indicating that the gains are unlikely to be explained by chance alone. Because the LLM does not receive cascading-failure outputs during inference, this pattern suggests that STRIDE/MITRE-derived semantics, together with topology and coupling information, encode signals correlated with the cascading-failure-based impact ranking.

Alignment with the fusion-based baseline (Figs.~\ref{fig:node_similarity} and~\ref{fig:path_similarity}) highlights the importance of combining threat intelligence with system-level impact. Mistral shows the strongest alignment, while LLaMA and Qwen rely more on structural cues. Despite having no direct access to cascading-failure outputs during inference, LLMs achieve meaningful agreement with fusion-based references, indicating that structured prompting captures key cyber-physical dependencies. Path-level similarity remains lower than node-level alignment, reflecting the greater difficulty of modeling precise transition dynamics.

The proposed framework reduces runtime by approximately 75--85\% compared to the fusion-based baseline (Table~\ref{tab:runtime}), demonstrating practical efficiency in small-scale settings. Overall, integrating structured threat modeling with LLM-based reasoning enables efficient identification of impact-aligned attack paths, with improvements over both topology-driven and structure-only baselines highlighting the importance of threat semantics.

\section{Conclusion and Future Work}
\label{sec:conclusion}
This paper presents a physics-grounded LLM framework for cyber-physical attack path identification, integrating STRIDE/MITRE threat modeling with system topology. In the IEEE 14-bus setting, the approach improves alignment with high-impact nodes, achieves meaningful agreement with fusion-based references, and reduces runtime by 75--85\%. These results suggest that structured LLM reasoning can serve as an efficient proxy for impact-aware analysis without requiring cascading-failure simulation during path generation.

Future work will focus on evaluating the framework's scalability on larger CPPS test cases and exploring the integration of physics-based signals, such as cascading-failure impact scores, into the LLM pipeline. This includes investigating fine-tuning on cascading-failure outcomes to improve alignment and path consistency.
\bibliography{ref}
\bibliographystyle{IEEEtran}
\end{document}